# Focusing of Parametric X-ray Radiation from a Bent Crystal

A.V. Shchagin

Kharkov Institute of Physics and Technology, Kharkov 61108, Ukraine

E-mail: shchagin@kipt.kharkov.ua

**Abstract**

An idea of producing a focused parametric X-ray radiation (PXR) is presented. The PXR is emitted by relativistic charged particles channeling along a bent crystal. Then the PXR emitted from the whole length of the bent crystal is going to the focus. Some properties of focused PXR are estimated for typical experimental conditions and possible applications are discussed. The experiment for observation of focused PXR is proposed.

## Introduction

A possibility of steering charged particles motion by a bent crystal has first been predicted in Ref. [1]. The steering can be realized for particles which are traveling through a long bent crystal in the channeling regime [2]. The channeling particles may be deflected, if the crystal is bent smoothly enough. Experiments on proton beam deflection, focusing and extraction [3] by bent crystals have been performed, e.g., in Serpukhov at 70 GeV [4, 5], CERN at 120 GeV [6] and at 450 GeV [7,8], Fermilab at 900 GeV [9]. The steering of positive particles at planar channeling is most effective (see, e.g., [7]). Bent crystals of about several centimeters in length are generally used in the experiments. These crystals seem to be convenient for production of focused parametric X-ray radiation (PXR) by channeling relativistic particles.

Experimental studies of PXR from relativistic electrons moving through a thin crystal have been performed since 1985 to gain an insight into the PXR properties and to develop a new source of quasimonochromatic polarized X-ray beam. The PXR has its maximum intensity in the vicinity of Bragg direction relative to the crystallographic plane (PXR reflection). The PXR reflection from crystals was observed and studied at electron beam energies ranging from several MeV to several GeV, and at PXR energies from a few keV to hundreds of keV. The validity of kinematical theory [10] for the description of PXR properties has been demonstrated in the most of related publications. General information about the nature, properties and investigations of the PXR, as well as references to original papers can be found, e.g., in reviews [11-14]. The experiment on observation of PXR from protons has been described in Ref. [15]. However, the PXR from particles channeling in a bent crystal has not been studied as yet, so far as we know.

Crystals usually have a number of crystallographic planes aligned at different directions. Therefore, relativistic particles emit a number of PXR reflections at Bragg directions relative to these crystallographic planes (see, e.g. [16]). In the present paper we will show a possibility of focusing the PXR generated in a bent crystal by channeling particles and consider some of its properties. The experiment for observation of focused PXR is also proposed.

**How to focus PXR**

Let us suppose that relativistic charged particles are moving in the channeling regime along a bent crystalline plate, as it is shown in Fig. 1. The bent crystal is cylindrical in shape, $R$ being the radius, and $f$ is the axis of the cylinder. The particles are channeling along the crystallographic planes denoted by the reciprocal lattice vector $\vec{g}$. In this paper we will consider PXR reflections from crystallographic planes aligned at $45°$ relative to the particle trajectory and perpendicular to the plane of Fig. 1a. These crystallographic planes are denoted by the reciprocal lattice vectors $\vec{g}_1$ and $\vec{g}_2$. The PXR reflections from these planes are going perpendicularly to the particle trajectory and the crystalline plate surface. The PXR reflections from $\vec{g}_1$ are going from the whole plate to the axis $f$, or, in other words, are focused. The PXR reflections from $\vec{g}_2$ are going in opposite directions, i.e., from the axis $f$. Thus, the PXR generated on the crystallographic planes $\vec{g}_1$ in the whole bent crystal of several centimeters in length will be collected (focused) at the axis $f$.

The properties of the PXR reflection generated at the right angle to the particle beam in thin crystals have been studied in several papers. The image of the whole PXR reflection at the right angle to the particle beam was first observed in Ref. [17], its polarization was considered in [18] and a detailed structure of the yield angular distribution was studied in [19]. The shape of angular distribution of the yield may be found, e.g., in figures of Refs. [17, 19, 14]. The shape of angular distribution of linear polarization directions in the PXR reflection at the right angle to the particle trajectory may be found, e.g., in figures of Refs. [18,14]. The angular distribution of the PXR yield in the reflection at the right angle to the particle trajectory has two maxima in the plane of Fig. 1b. Therefore the PXR will be focused at two points $F_1$ and $F_2$ at the axis $f$ at an angular distance of about $2\gamma_{eff}^{-1}$ one from another. Here, $\gamma_{eff}$ is the effective relativistic factor in the medium [14], $\gamma_{eff}^{-1} = \sqrt{\gamma^{-2} + |\chi_0|}$, where $\gamma$ is the relativistic factor of incident particles, $\chi_0$ is the dielectric susceptibility of the medium. The PXR is linearly polarized in the vicinity of points $F_1$ and $F_2$. The linear polarization direction is practically in the plane of Fig. 1b. Thus, the focused PXR is linearly polarized in the plane of Fig. 1b.

Apart from the PXR emission, the particles moving through the crystalline plate ionize crystal atoms and thus make them emit the characteristic X-ray radiation (CXR). The CXR is isotropic and not focused, but it exists at points $F_1$ and $F_2$, too.

## Estimations of some properties of focused PXR

To understand the main features of focused PXR, let us estimate them for typical experimental conditions. As an example, we performed calculations for 450 GeV protons channeling along the crystallographic plane (110) with the reciprocal lattice vector $\vec{g} = <\bar{1}\bar{1}0>$ (see Fig. 1) in the Si single-crystal plate, 0.5 mm in thickness and $l_{cryst} = 5$ cm in length, with the curvature radius $R = 5$ m. The PXR reflections under consideration are generated at crystallographic planes parallel to the ones denoted by the reciprocal lattice vectors $\vec{g}_1 = <100>$ and $\vec{g}_2 = <0\bar{1}0>$. The calculations were performed using formulae from Refs. [12, 14, 20] derived within the framework of the Ter-Mikaelian theory [10]. The focused PXR reflections are going to the points $F_1$ and $F_2$ from the crystallographic planes having a nonzero structure factor parallel to (100), these are (400), (800), (12 00). The PXR reflections from higher-index planes have substantially lower yields. The PXR reflections from similar crystallographic planes, that are parallel to $(0\bar{1}0)$, are going in opposite directions from the axis $f$. The particular case of channeling radiation diffraction [21] is not considered in the present estimations.

The crystallographic axis $<\bar{1}10>$ is aligned along the proton beam in the above-considered example. To avoid axial channeling in the experiment, one needs to provide misalignment of the crystal lattice in the (110) plane for a small angle, which is somewhat larger than the critical channeling angle. The critical channeling angle for the Si single crystal at a proton energy of 450 GeV is about $10^{-5}$ radian [2]. The main features of PXR emission, the positions of points $F_1$, $F_2$ and the properties of focused PXR will change insignificantly at such a small misalignment.

The PXR properties are functions of the relativistic factor of the incident particle and the absolute value of particle charge [10,12,20]. Therefore, the results of present calculations might be valid for 245 MeV positrons and electrons, which have the same relativistic factor. However, the motion of these particles in a long crystal can differ from the motion of protons. For example, a significant part of high-energy protons can move through a long bent crystal in the channeling regime [2,4-9]. But electrons strongly scatter at channeling. So far as we know, there have been no publications about experimental research on the channeling of positrons in bent crystals and on generation of PXR by positrons in any crystals.

Besides, the yield of CXR at 1.74 keV due to K-shell ionization of Si atoms by 450 GeV protons was estimated in accordance with the recommendations from Ref. [22]. In the calculations, the K-shell ionization cross section for Si atoms was taken to be 1943 barn. This cross section value was calculated for 245 MeV electrons having the same relativistic factor as that of 450 GeV protons. The effect of channeling was not taken into account in the present estimations of CXR yield, as it is not expected to be very significant at high particle energies (see Refs. in [22]).

The calculations were performed for a uniform distribution of protons in the crystalline plate thickness. The radiation excited by channeling protons is going to points $F_1$ and $F_2$ across the plate, and is attenuated by the Si crystal. The attenuation of radiation in the plate was taken into account in the calculations.

## Results and discussion

The results of calculations are presented in Table 1. The spectrum of X-ray radiation at points $F_1$ and $F_2$ has four spectral peaks in the 1.7 – 19.4 keV energy range, this being convenient for registration by standard X-ray spectrometric detectors. The peak at 1.74 keV is due to a non-focused non-polarized CXR, which is produced mainly by protons moving in a 13.3 $\mu m$ layer at a concave side of the plate. This layer thickness is determined by X-ray attenuation in a Si crystal (see Table 1). Thus, only a little part of the total number of protons passing through the plate produces the CXR that reaches the points $F_1$ and $F_2$. The other spectral peaks are due to a focused polarized PXR. The spectral peaks of focused PXR at $E = 6.46$, 12.91, 19.37 keV are produced mainly by protons moving in the 37 $\mu m$ layer at a concave side of the plate, in the 270 $\mu m$ layer at the same side of the plate, and in almost the whole crystalline plate thickness respectively.

The radiation sources for different spectral peaks are specifically distributed in the crystal thickness. Therefore, the measurements and analysis of relative intensities of spectral peaks would permit the estimation of beam proton distribution in the crystal thickness. This can be done for the whole crystalline plate or for its separate part, provided that other parts of the plate are screened, and the X-ray detector can see only this separate part of the crystal. The CXR from the 13.3 $\mu m$ layer at a convex side of the bent crystal may be registered by an X-ray detector installed opposite to a convex side of the bent crystal, as the CXR is isotropic and non-focused.

Angular divergence (convergence) of focused PXR may be controlled by changing the radius of curvature $R$ of the bent crystal. Thus, we have an X-ray source of several centimeters in size, with a provided smooth tuning of the divergence (convergence). This monochromatic source of polarized X-rays with a possibility of smooth tuning of divergence may be useful for calibration of large-aperture X-ray equipment, for example, X-ray space telescopes [16]. The divergence (convergence) of PXR reflections from channeling particles is $\frac{l_{cryst}}{R}$ at the plane of Fig. 1a.

The focused PXR generated by channeling positive relativistic particles in a long crystal can provide a natural PXR spectral peak of extremely narrow width. The natural spectral peak width of focused PXR may be of microelectronvolts range at PXR spectral peak energy of about 10 keV [23].

The experiment on observation of focused PXR may be performed at a facility with a crystal bent for steering the proton beam having the channeling particles intensity of about or above $10^7 \frac{p}{s}$. It can be seen from Table 1 that all spectral peaks have comparable intensities, and thus, can be measured simultaneously by a spectrometric X-ray detector. The X-ray detector, being about 1 cm$^2$ square, should be installed at point $F_1$ or $F_2$ at the axis $f$. The PXR from non-channeled particles can be rejected by registering X-rays in coincidence with the particles that have passed through the bent crystal in the channeling regime. There should be vacuum connection between the bent crystal and the X-ray detector in order to observe soft X-rays, because of their attenuation in air. X-rays of energies 1.74 and 6.46 will be absorbed practically completely over a distance of 5 m in air. However, observation of higher-energy focused PXR is possible in air. The X-rays of energies 12.91 and 19.37 keV will be attenuated by factors of about 0.23 and 0.60, respectively, at 5 m in air. The energies of all spectral peaks are practically independent of incident relativistic particle energy. Yet, their intensities can increase (decrease) with an increasing (decreasing) incident proton energy [12,20].

## Acknowledgements

The author is thankful to E.N. Tsyganov for discussions during the meeting, [3,24] and more recently, to X. Artru and I. Endo and V.V. Sotnikov. This work was presented as an oral paper at Workshop [25]. The paper became possible partially due to grant STCU 1030.

**Table 1.**

Table 1. Properties of CXR and focused PXR from 450 GeV protons channeling in a bent Si single-crystal plate, 0.5 mm in thickness and 5 cm in length, along the crystallographic plane (110). The radius of crystal curvature is 5 m. The origin of radiation (CXR or focused PXR) and polarization properties of the radiation are given in the first and second columns of the table, where $E$ is the energy of spectral peaks of radiation, $\gamma_{eff}^{-1}$ is the inverse effective relativistic factor (for comparison, the inverse relativistic factor of incident 450 GeV protons is $\gamma^{-1} = 2.08 \cdot 10^{-3}$), $\Delta F$ is the distance between points $F_1$ and $F_2$, $T_e$ is the e-fold attenuation length of radiation having the energy $E$ in a Si single crystal, $I$ is the number of quanta per proton per $cm^2$ at points $F_1$ and $F_2$ with X-ray attenuation in the crystal taken into account for the total number of channeling protons randomly distributed in the plate thickness.

| Radiation | Polari-zation | $E, keV$ | $\gamma_{eff}^{-1}$ | $\Delta F, mm$ | $T_e$ in Si, $\mu m$ | $I, \dfrac{quanta}{cm^2 \cdot p^+}$ |
|---|---|---|---|---|---|---|
| CXR Si | No | 1.74 | | | 13.3 | $1.93 \cdot 10^{-7}$ |
| PXR(400) | Linear | 6.46 | $5.24 \cdot 10^{-3}$ | 52.4 | 37 | $6.90 \cdot 10^{-7}$ |
| PXR(800) | Linear | 12.91 | $3.18 \cdot 10^{-3}$ | 31.8 | 270 | $3.32 \cdot 10^{-7}$ |
| PXR(12 00) | Linear | 19.37 | $2.63 \cdot 10^{-3}$ | 26.3 | 865 | $1.28 \cdot 10^{-7}$ |

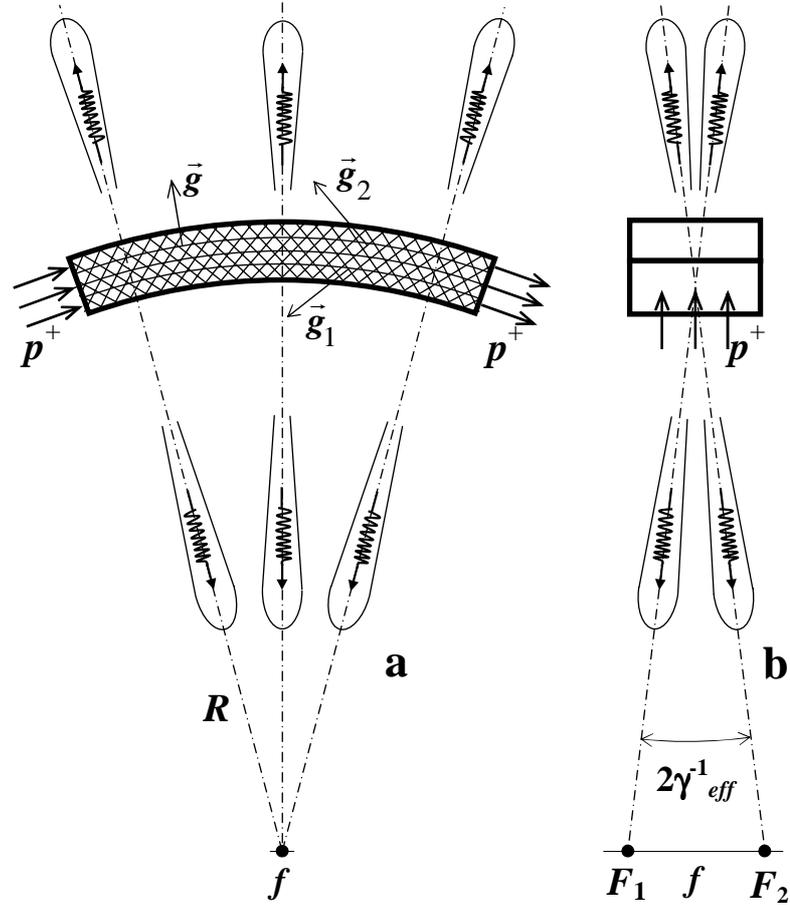

Fig. 1. Production of focused PXR by particles channeling in a bent crystalline plate. The side and front views are shown in Figs. 1a and 1b, respectively. The particle beam is indicated by arrows $p^+$. The particles are channeling along the crystallographic planes denoted by the reciprocal lattice vector $\vec{g}$. The single-crystal plate is cylindrical in shape with a radius of curvature $R$ around the axis $f$. The angular distributions of the yields in the PXR reflections are shown schematically for three points at the plate. Wavy lines with arrows show the radiation propagation direction in the PXR reflections. PXR reflections from crystallographic planes denoted by the reciprocal lattice vector $\vec{g}_1$ are focused at points $F_1$ and $F_2$ at the axis $f$. PXR reflections from crystallographic planes denoted by the reciprocal lattice vector $\vec{g}_2$ are going in opposite directions. The PXR focused at points $F_1$ and $F_2$ is linearly polarized practically in the plane of Fig. 1b.